\documentclass[12pt]{article}
\usepackage{setspace}
\usepackage{authblk}
\usepackage{pgfplotstable,filecontents}
\RequirePackage[OT1]{fontenc}
\usepackage{microtype}
\usepackage{caption} 

\RequirePackage[round,sort&compress]{natbib}
\RequirePackage[colorlinks,citecolor=blue,urlcolor=blue]{hyperref}
\RequirePackage{hypernat}

\usepackage{amsmath,amsfonts,amssymb}
\usepackage{graphicx} 
\usepackage{booktabs} 
\usepackage{paralist}
\usepackage{amsbsy}
\usepackage{JASA_manu}

\usepackage[mathscr]{eucal}

\usepackage{bm}

\usepackage{xspace}

\usepackage{algorithm,algorithmicx}
\usepackage{algpseudocode}

\mathchardef\mhyphen="2D

\ifpdf
\newcommand{\Sec}[1]{\hyperref[sec:#1]{\S\ref*{sec:#1}}} 
\newcommand{\App}[1]{\hyperref[sec:#1]{Appendix~\ref*{sec:#1}}} 
\newcommand{\Eqn}[1]{\hyperref[eq:#1]{{\rm (\ref*{eq:#1})}}} 
\newcommand{\Part}[1]{\hyperref[part:#1]{(\ref*{part:#1})}} 
\newcommand{\Fig}[1]{\hyperref[fig:#1]{Figure~\ref*{fig:#1}}} 
\newcommand{\Tab}[1]{\hyperref[tab:#1]{Table~\ref*{tab:#1}}} 
\newcommand{\Thm}[1]{\hyperref[thm:#1]{Theorem~\ref*{thm:#1}}} 
\newcommand{\Lem}[1]{\hyperref[lem:#1]{Lemma~\ref*{lem:#1}}} 
\newcommand{\Prop}[1]{\hyperref[prop:#1]{Proposition~\ref*{prop:#1}}} 
\newcommand{\Cor}[1]{\hyperref[cor:#1]{Corollary~\ref*{cor:#1}}} 
\newcommand{\Def}[1]{\hyperref[def:#1]{Definition~\ref*{def:#1}}} 
\newcommand{\Alg}[1]{\hyperref[alg:#1]{Algorithm~\ref*{alg:#1}}} 
\newcommand{\Ex}[1]{\hyperref[ex:#1]{Example~\ref*{ex:#1}}} 
\newcommand{\As}[1]{\hyperref[as:#1]{Assumption~{\rm\ref*{as:#1}}}} 
\newcommand{\Reg}[1]{\hyperref[as:#1]{Condition~\ref*{reg:#1}}} 
\newcommand{\AlgLine}[2]{\hyperref[alg:#1]{line~\ref*{line:#2} of Algorithm~\ref*{alg:#1}}}
\newcommand{\AlgLines}[3]{\hyperref[alg:#1]{lines~\ref*{line:#2}--\ref*{line:#3} of Algorithm~\ref*{alg:#1}}}
\else
\newcommand{\Sec}[1]{{\S\ref{sec:#1}}} 
\newcommand{\App}[1]{{Appendix~\ref{sec:#1}}} 
\newcommand{\Eqn}[1]{{(\ref{eq:#1})}} 
\newcommand{\Part}[1]{{(\ref{part:#1})}} 
\newcommand{\Fig}[1]{{Figure~\ref{fig:#1}}} 
\newcommand{\Tab}[1]{{Table~\ref{tab:#1}}} 
\newcommand{\Thm}[1]{{Theorem~\ref{thm:#1}}} 
\newcommand{\Lem}[1]{{Lemma~\ref{lem:#1}}} 
\newcommand{\Prop}[1]{{Proposition~\ref{prop:#1}}} 
\newcommand{\Cor}[1]{{Corollary~\ref{cor:#1}}} 
\newcommand{\Def}[1]{{Definition~\ref{def:#1}}} 
\newcommand{\Alg}[1]{{Algorithm~\ref{alg:#1}}} 
\newcommand{\Ex}[1]{{Example~\ref{ex:#1}}} 
\newcommand{\Reg}[1]{{R~\ref*{reg:#1}}} 
\fi

\newcommand{\Real}{\mathbb{R}}

\newcommand{\Tra}{^{\sf T}} 

\newcommand{\V}[1]{{\bm{\mathbf{\MakeLowercase{#1}}}}} 
\newcommand{\VE}[2]{\MakeLowercase{#1}_{#2}} 
\newcommand{\Vtilde}[1]{{\bm{\tilde \mathbf{\MakeLowercase{#1}}}}} 
\newcommand{\Vn}[2]{\V{#1}^{(#2)}} 

\newcommand{\M}[1]{{\bm{\mathbf{\MakeUppercase{#1}}}}} 
\newcommand{\ME}[2]{\MakeLowercase{#1}_{#2}} 
\newcommand{\Mtilde}[1]{{\bm{\tilde \mathbf{\MakeUppercase{#1}}}}} 
\newcommand{\Mn}[2]{\M{#1}^{(#2)}} 

\begin{document}
\title{{\bf $k$-POD}  \\ A Method for $k$-Means Clustering of Missing Data}
\author{Jocelyn T. Chi}
\author{Eric C. Chi}
\author{Richard G. Baraniuk}
\affil{Department of Electrical and Computer Engineering \\
Rice University \\
Houston, TX 77005}
\date{}
\maketitle

\mbox{}
\vspace*{1.75in}
\begin{center}
\textbf{Author's Footnote:}
\end{center}
Jocelyn T. Chi is Ph.D. Student, Department of Statistics, North Carolina State University, Raleigh NC 27695 (E-mail: jtchi@ncsu.edu);
Eric C. Chi is Assistant Professor, Department of Statistics, North Carolina State University, Raleigh NC 27695 (E-mail: ecchi@ncsu.edu); and
Richard G. Baraniuk is Professor, Department of Electrical and Computer Engineering, Rice University, Houston TX 77005 (E-mail: richb@rice.edu).

\newpage

\begin{abstract}
The $k$-means algorithm is often used in clustering applications but its usage requires a complete data matrix.  Missing data, however, is common in many applications.  Mainstream approaches to clustering missing data reduce the missing data problem to a complete data formulation through either deletion or imputation but these solutions may incur significant costs.  Our $k$-POD method presents a simple extension of $k$-means clustering for missing data that works even when the missingness mechanism is unknown, when external information is unavailable, and when there is significant missingness in the data.
\end{abstract}

\vspace*{.3in}


\newpage

\section{Introduction}

The clustering problem is ubiquitous in exploratory data analysis.  Given a collection of objects, we wish to group them so that members within the same group are similar and members in different groups are dissimilar. Although the idea is simple, the volume of literature dedicated to it \citep{Gor1999,Har1975,KauRou1990,Mir1996,WuWun2009} testifies to its challenges; there is no single clustering method that works universally well in all contexts.

Despite the plethora of clustering methods designed to address the diversity of context-specific challenges, far less attention has been given to the clustering problem when data are missing, or partially observed.  Yet, missing data is common in many clustering applications.  In astronomy, for instance, imaging sensors have limited sensitivity and may fail to detect light intensities below a minimal threshold \citep{WagLai2005}, frustrating the clustering of celestial bodies.  Similarly, survey non-response presents a quandary in clustering respondents in social surveys \citep{BriKal1996}.  To reconcile the reality of missing data and the fact that clustering methods typically require complete data, practitioners often convert a partially observed dataset to a completely observed one through one of two strategies: deletion or imputation \citep{WagLai2005, dixon}.

Deletion achieves complete data through subtraction; it removes variables containing missing entries.  Despite its simplicity, deletion can sometimes be adequate for clustering purposes when relatively few variables are missing entries. On the other hand, deletion is inadvisable when a substantial fraction of variables are affected or when the missingness is meaningful to the clusterings. 

Imputation achieves complete data through addition; it fills in missing entries with plausible estimates of their values.  Unlike the simple deletion procedure, imputation methods can be substantially more complicated, since effective imputation requires information on the joint distribution of the missingness patterns and the data.  Although probabilistic models can work extremely well when the assumptions are correct, they often entail time-consuming computations.  Furthermore, although we can only hypothesize models for the missingness in most cases, imputations rely on correct ascertainment of the distributional assumptions on the missingness patterns.  When the missingness depends on the unobserved data, imputation is both laborious and uncertain since modeling error cannot be determined without external validation~\citep{mice}.  If we are primarily interested in clustering the observations --- and are uninterested in estimating values for the missing entries --- then imputation is, at best, a computationally expensive pre-processing step.  At worst, if the distributional assumptions are substantially incorrect, imputation can also lead to poor clusterings.  Throughout this paper, we use the terms {\em imputation} and {\em impute} to refer to state-of-the art probabilistic imputation methods and their procedures.

In an effort to avoid imputation and to mitigate the hazards of deletion, some have also considered augmented deletion.  For example, \cite{WagLai2005} proposed a method for augmenting $k$-means clustering on deleted data with ``soft constraints" for astronomy applications.  The method augments classical $k$-means clustering on deleted data with weighted penalties consisting of a partial measure of dissimilarity between observations.  This results in a tuning parameter for each variable containing missing entries based on the known relative importance of the variable in clustering the observations.  However, there are no guidelines on how to select the tuning parameters when this relative importance remains unknown.  While augmented deletion may offer a step in the right direction for some applications, we seek a general alternative that requires neither tuning parameters nor additional information and assumptions.

Faced with the dilemma of wasted data or time-consuming and potentially erroneous imputations, we present a new alternative: $k$-POD, a novel method of $k$-means clustering on partially observed data.  The $k$-POD method employs a majorization-minimization (MM) algorithm~\citep{BecYanLan1997,LanHunYan2000} to identify a clustering that is in accord with the observed data.  By bypassing the completely observed data formulation, $k$-POD retains all information in the data and avoids committing to distributional assumptions on the missingness patterns.

The $k$-POD method distinguishes itself from current approaches to $k$-means clustering of missing data in three ways: i) it is simple, ii) it is fast, and iii) it performs reliably even at large overall percentages of missingness.  Since it does not require assumptions on the missingness pattern and utilizes a simple, tuning-free descent algorithm, it simply works ``out of the box."  

There are two fundamental differences between $k$-POD and approaches to clustering missing data that utilize state-of-the-art imputation methods.  First, these imputation-clustering approaches work well when they can identify plausible values for the missing data.  In practice, however, there is no way to verify the accuracy of the imputations.
In contrast, $k$-POD minimizes the sum of the squared differences between the data and the resulting clusterings over the observed entries only.  By focusing on differences over the observed data, $k$-POD remains unhindered by the need for accurate imputations.

A key point of novelty in the $k$-POD method is the combination of a formulation that is common to matrix completion problems with a descent algorithm in the MM framework to produce clusterings that agree with the observed data.  Remarkably, the method works accurately and efficiently --- without the setup and computationally expensive work typically required of probabilistic imputation methods.

Our numerical experiments below present a survey of clustering partially observed data under different missingness patterns.  While current approaches yield good results at low levels of missingness, they fail to produce results at larger overall missingness percentages or require prohibitively long computation times.  Even on some modestly sized datasets, we show that imputation can lead to impractical pre-processing times.  In contrast, $k$-POD produces accurate clusterings, regardless of the missingness mechanism, even at large percentages of overall missingness, and within reasonable time.

Finally, we note that there are some extensions of mixture model clustering to handle missing data \citep{GhaJor1994, HunJor2003, LinLeeHo2006}. The accompanying expectation-maximization (EM) algorithms for estimating the parameters of these models bear some similarity to $k$-POD. This is not surprising, given that $k$-means can be seen as an approximation to the classic EM clustering algorithm \citep{KulJor2012}. Determining the precise relationship between $k$-POD and the EM algorithms for fitting these mixture models, however, is beyond the scope of this paper. While more sophisticated --- and computationally demanding --- mixture models can potentially identify more complicated cluster structures when data is missing, our aim in this work is to show how to extend the simple, and computationally efficient, $k$-means algorithm to handle missing data in a way that maintains its simplicity and efficiency.
To the best of our knowledge, our $k$-POD method for $k$-means clustering of missing data has not been proposed before in the literature.

%
\section{Preview}
\label{sec:intuition}

To illustrate the intuition behind the $k$-POD method for clustering missing data, we begin by examining the $k$-means algorithm.  Essentially, $k$-means works by estimating cluster centroids based on cluster assignments.  Its underlying assumption is that each observation --- or row in a data matrix --- is a noisy realization of a cluster centroid.  This key premise affords the following intuitive approach to $k$-means clustering with missing data.

Suppose that we have a matrix ${\bf Y}$ of partially observed data, in which each observation is a noisy instance of a known cluster centroid.  If we also know the cluster membership of each observation, then a very reasonable thing to do would be to estimate the missing entries in ${\bf Y}$ with the corresponding entries from the relevant centroid.  Once we have this complete data matrix, we can use $k$-means to cluster the observations again.  If the cluster assignments and centroids change, we can then update our estimates for the missing entries in ${\bf Y}$ with the corresponding entries from the new cluster assignments and centroids.

In fact, this is the basic idea behind how $k$-POD works.  We can make this intuition mathematically rigorous because the procedure can be formulated as an MM algorithm for minimizing the objective function of a missing data version of the $k$-means problem.  

This idea also illustrates a fundamental difference between the $k$-POD method and approaches to clustering missing data that rely on imputation methods.  All probabilistic imputation approaches have assumptions, but those assumptions are not based on the clusterings.  In contrast, the underlying assumption in $k$-POD is that each observation is a noisy instantiation of a cluster centroid.  By combining this premise with the MM framework, $k$-POD ensures that each subsequent estimate for the missing values improves the objective function.

In the following two sections, we describe the mathematical formulations for the $k$-means problem and an MM algorithm for a missing data version of the $k$-means clustering problem.

%
\section{The \lowercase{$k$}-means problem}
\label{sec:kmeans}

Given a data matrix $\M{Y} \in \Real^{n \times p}$ of $n$ observations and $p$ features, our task is to cluster the $n$ observations into $k$ clusters. We first set some notation in order to pose the problem.  Let $\mathcal{C} = \{C_1, \ldots, C_k\}$ denote a partition of the $n$ observations into $k$ clusters, namely the sets in $\mathcal{C}$ are disjoint and $\cup_{i=1}^k C_i = \{1, \ldots, n\}.$  Let $\M{B} \in \Real^{k \times p}$ denote a matrix whose rows are the centroids of the clusters, namely $\M{B}\Tra = \begin{pmatrix}\V{b}_1\Tra & \cdots \V{b}_k\Tra \end{pmatrix}$, where $\V{b}_i \in \Real^{p}$ is the centroid associated with the $i$th partition $C_i$.

The $k$-means problem seeks the partition and centroid values that minimize the following sum of squared residuals\begin{equation}
\label{eq:kmeans0}
\min_{\mathcal{C},\M{B}} \sum_{i=1}^k \sum_{j \in C_i} \lVert \V{y}_j - \V{b}_i \rVert_2^2,
\end{equation}
where $\V{y}_j$ is the $j$th row of $\M{Y}$.  This is an NP-hard problem \citep{AloDesHan2009,DasFre2009} and hence is typically solved using a greedy alternating minimization algorithm, the most popular of which is Lloyd's method \citep{For1965,Mac1967,Llo1982,HarWon1979}.

We rewrite \Eqn{kmeans0} in terms of the matrices $\M{Y}$ and $\M{B}$ to set the stage for formulating a missing data version. Recall that the Frobenius norm of a matrix is the square root of the sum of the squares of its entries, $\lVert \M{A} \rVert_{\text{F}}^2 = \sum_{ij} \VE{a}{ij}^2$. We can rewrite the minimization in \Eqn{kmeans0} as
\begin{equation}
\label{eq:kmeans1}
\min_{\M{A} \in H, \M{B}} \lVert \M{Y} - \M{A}\M{B} \rVert_{\text{F}}^2,
\end{equation}
where we parametrize the optimization over the partitions $\mathcal{C}$ by optimization of the membership matrices defined by the set $H = \{\M{A} \in \{0,1\}^{n \times k} : \M{A}\V{1} = \V{1} \}$. The binary matrix $\M{A}$ encodes the cluster memberships. The $i,j$th entry $\ME{a}{ij}$ of the matrix $\M{A} \in H$ is $1$ if $i \in C_j$ and zero otherwise.

In words, the $j$th row of $\M{A}$ is the transpose of the standard basis vector $\V{e}_i \in \Real^k$ if and only if the $j$th observation has been assigned to the $i$th cluster. Thus, the condition that $\M{A}\V{1} = \V{1}$ encodes the requirement that every observation is assigned to one, and only one, partition. 

We now formulate the missing data version of \Eqn{kmeans1}. Let $\Omega \subseteq \{1, \ldots, n\} \times \{1, \ldots, p\}$ be a subset of the indices that correspond to the observed entries. The projection operator of $n \times p$ matrices onto a index set $\Omega$ is given by
\begin{align*}
[P_\Omega(\M{Y})]_{ij} = & \begin{cases}
\ME{Y}{ij} & \text{if $(i,j) \in \Omega$} \\
0 & \text{if $(i,j) \in \Omega^c$.}
\end{cases}
\end{align*}
We propose that a natural formulation of a missing data version of the $k$-means problem \Eqn{kmeans1} seeks to solve the following problem:
\begin{equation}
\label{eq:kmeans_missing}
\min_{\M{A} \in H, \M{B}} \lVert P_{\Omega}(\M{Y}) - P_{\Omega}(\M{A}\M{B}) \rVert_{\text{F}}^2,
\end{equation}
namely we seek the factorization $\M{A}\M{B}$ of the data matrix $\M{Y}$ that minimizes the sum of squared errors over the observed entries $\Omega$. Although the discrepancy between the data matrix $\M{Y}$ and a model matrix $\M{A}\M{B}$ over the observed entries $\Omega$ is a key quantity in matrix completion problems \citep{CanRec2009,CaiCanShe2010,MazHasTib2010}, we emphasize that our primary concern is clustering and not imputation. Employing a matrix completion formulation of the incomplete $k$-means problem makes our MM solution readily apparent.

\section{Majorization-Minimization Algorithm}
\label{sec:mm}

We now develop a simple MM algorithm for solving the minimization in \Eqn{kmeans_missing}. The basic strategy behind an MM algorithm is to convert a hard optimization problem into a sequence of simpler ones. The MM principle requires majorizing the objective function $f(\V{u})$ by a surrogate function $g(\V{u} \mid \Vtilde{u})$ anchored at $\Vtilde{u}$.  Majorization is a combination of the tangency condition $g(\V{u} \mid {\bf \tilde{u}}) =  f( {\bf \tilde{u}})$ and the domination condition $g(\V{u} \mid \Vtilde{u})  \geq f(\V{u})$ for all $\V{u} \in \Real^n$.  The associated MM algorithm is defined by the iterates $\Vn{u}{m+1} := \underset{\V{u}}{\arg \min}\; g(\V{u} \mid \Vn{u}{m})$. It is straightforward to verify that the MM iterates generate a descent algorithm driving the objective function downhill, namely that $f(\Vn{u}{m+1}) \leq f(\Vn{u}{m})$ for all $m$. Note that we still obtain the monotonicity property even if we do not exactly minimize the majorization. That will be the case when $k$-means is applied to our majorization.

Returning to our original problem,  we observe that the following function of $(\M{A},\M{B})$ is non-negative
\begin{equation*}
\lVert P_{\Omega^c}(\M{A}\M{B}) - P_{\Omega^c}(\Mtilde{A}\Mtilde{B}) \rVert_{\text{F}}^2 \geq 0,
\end{equation*}
and the inequality becomes equality when $\M{A}\M{B} = \Mtilde{A}\Mtilde{B}$, which occurs when $(\M{A},\M{B}) = (\Mtilde{A},\Mtilde{B})$. Adding the above non-negative function to the missing data version of the $k$-means objective gives us the following function
\begin{align*}
g(\M{A},\M{B} \mid \Mtilde{A},\Mtilde{B}) & =  \lVert P_{\Omega}(\M{Y}) - P_{\Omega}(\M{A}\M{B}) \rVert_{\text{F}}^2
+ \lVert P_{\Omega^c}(\M{A}\M{B}) - P_{\Omega^c}(\Mtilde{A}\Mtilde{B}) \rVert_{\text{F}}^2 \\
& = \lVert \Mtilde{Y} - \M{A}\M{B} \rVert_{\text{F}}^2,
\end{align*}
where $\Mtilde{Y} = \mathcal{P}_{\Omega}(\M{Y}) + \mathcal{P}_{\Omega^c}(\Mtilde{A}\Mtilde{B})$. 
The function $g(\M{A},\M{B} \mid \Mtilde{A},\Mtilde{B})$ majorizes $\lVert P_{\Omega}(\M{Y}) - P_{\Omega}(\M{A}\M{B}) \rVert_{\text{F}}^2$ at the point $(\Mtilde{A},\Mtilde{B})$, since $g$ satisfies the domination condition
\begin{align*}
g(\M{A},\M{B} \mid \Mtilde{A},\Mtilde{B}) & \geq \lVert P_{\Omega}(\M{Y}) - P_{\Omega}(\M{A}\M{B}) \rVert_{\text{F}}^2
\end{align*}
for all $(\M{A},\M{B})$ and the tangency condition
\begin{align*}
g(\Mtilde{A},\Mtilde{B} \mid \Mtilde{A},\Mtilde{B}) & = \lVert P_{\Omega}(\M{Y}) - P_{\Omega}(\Mtilde{A}\Mtilde{B}) \rVert_{\text{F}}^2.
\end{align*}

Now consider the MM update rule given the $m$th iterate $(\Mn{A}{m},\Mn{B}{m})$. Let $(\Mn{A}{m+1},\Mn{B}{m+1})$ be the output of applying $k$-means clustering to $\Mn{Y}{m} = \mathcal{P}_{\Omega}(\M{Y}) + \mathcal{P}_{\Omega^c}(\Mn{A}{m}\Mn{B}{m})$.
First, the $k$-means clustering algorithm is monotonic, namely
\begin{align}
\label{eq:kmeans}
\lVert \Mn{Y}{m} - \Mn{A}{m}\Mn{B}{m} \rVert_{\text{F}}^2 & \geq \lVert \Mn{Y}{m} - \Mn{A}{m+1}\Mn{B}{m+1} \rVert_{\text{F}}^2.
\end{align}
Second, since $g$ is a majorization, we also know that
\begin{align}
\label{eq:tangency}
\lVert P_{\Omega}(\M{Y}) - P_{\Omega}(\Mn{A}{m}\Mn{B}{m}) \rVert_{\text{F}}^2 & = \lVert \Mn{Y}{m} - \Mn{A}{m}\Mn{B}{m} \rVert_{\text{F}}^2
\end{align}
and
\begin{align}
\label{eq:domination}
\lVert \Mn{Y}{m} - \Mn{A}{m+1}\Mn{B}{m+1} \rVert_{\text{F}}^2 & \geq \lVert P_{\Omega}(\M{Y}) - P_{\Omega}(\Mn{A}{m+1}\Mn{B}{m+1})\rVert_{\text{F}}^2.
\end{align}
Combining the equations in \Eqn{kmeans}, \Eqn{tangency}, and \Eqn{domination}, we arrive at the conclusion that $k$-POD drives the missing data objective downhill
\begin{align*}
\lVert P_{\Omega}(\M{Y}) - P_{\Omega}(\Mn{A}{m}\Mn{B}{m}) \rVert_{\text{F}}^2 & \geq \lVert P_{\Omega}(\M{Y}) - P_{\Omega}(\Mn{A}{m+1}\Mn{B}{m+1})\rVert_{\text{F}}^2.
\end{align*}

The $k$-POD algorithm is summarized in \Alg{MM}.  To obtain ${\bf A}^{0}$ and ${\bf B}^{0}$, we begin with a data matrix of partially observed data {\bf Y} and use a computationally inexpensive method to fill-in the unobserved entries in ${\bf Y}$.  In the numerical experiments below, we simply use column-wise means for this initialization.  We then employ $k$-means clustering to obtain initial cluster assignments and record the resulting cluster membership and centroid for each observation in ${\bf A}^{0}$ and ${\bf B}^{0}$, respectively.

The basic $k$-means algorithm is known to be sensitive to initialization, and solutions using random initializations can be arbitrarily bad with respect to an optimal solution. Consequently, we employ the {\tt k-means++} algorithm to choose starting points that are guaranteed to give clusterings within a factor of $\mathcal{O}(\log k)$ multiplicative error of the optimal $k$-means solution \citep{ArtVas2007}.

\begin{algorithm}[t]
  \caption{$k$-POD}
  \label{alg:MM}
\begin{algorithmic}[1]
\State Initialize $(\Mn{A}{0},\Mn{B}{0})$
\Repeat
\State $\Mn{Y}{m} \gets \mathcal{P}_{\Omega}(\M{Y}) + \mathcal{P}_{\Omega^c}(\Mn{A}{m}\Mn{B}{m})$
\Comment{Fill-in unobserved entries}
\State $(\Mn{A}{m+1},\Mn{B}{m+1}) \gets \text{$k$-means}(\Mn{Y}{m})$
\Comment{Update clustering}
\Until{convergence}
\end{algorithmic}
\end{algorithm}

In words, $k$-POD works as follows.  After initialization of ${\bf A}$ and ${\bf B}$ as described, repeat the following until convergence of the $k$-means algorithm.  First, update the unobserved portion of ${\bf Y}$ with the corresponding entries in ${\bf A}{\bf B}$.  Then, perform $k$-means clustering on the updated ${\bf Y}$ to obtain new cluster assignments and centroids.  Finally, use the new clustering result to update ${\bf A}$ and ${\bf B}$ as before.

Although each iteration utilizes an inexpensive fill-in step in the update procedure, we emphasize that the $k$-POD method differs from current state-of-the-art probabilistic imputation methods for clustering missing data in that it requires no assumptions on the missingness patterns, it is not concerned with the quality of the imputed data, and it produces clustering results that agree with the observed data.

\section{Numerical Experiments}
\label{sec:numericalcomparisons}

We compare $k$-POD to $k$-means clustering after imputation and deletion approaches under the three canonical missingness patterns or mechanisms.  In this section, we describe the missingness mechanisms, the comparison methods, the data, and the experimental setup.  

\subsection{Mechanisms of Missingness}
\label{sec:missingnessmechanisms}

We adopt the notation in \cite{littlerubin} to describe the mechanisms of missingness.  Let $\M{Y} \in \mathbb{R}^{n \times p}$ be a complete data matrix generated according to some parametric distribution indexed by the parameter $\theta$.  Let $\M{Y}_{\text{obs}} \in \mathbb{R}^{n \times p}$ denote the observed portion of $\M{Y}$, and let $\M{M} \in \mathbb{R}^{n \times p}$ be the indicator matrix with entries $\ME{m}{ij} = 1$ when $\ME{y}{ij}$ is missing and $\ME{m}{ij} = 0$ otherwise. Typically, we also assume that the missingness pattern depends on an unknown parameter $\phi$. Thus, a probabilistic model for the missingness pattern that led to $\M{Y}_{\text{obs}}$ is encoded in the conditional distribution of $\M{M}$ given $\M{Y}$ and the unknown parameter $\phi$, namely $f(\M{M} \mid \M{Y}, \phi)$. 

We write the joint conditional distribution of $\M{Y}$ and $\M{M}$ as 
\begin{align*}
f(\M{Y}, \M{M} \mid \theta, \phi)  & =  f(\M{M} \mid \M{Y}, \phi) f(\M{Y} \mid \theta).
\end{align*}
In this model, the data $\M{Y}$ is first generated according to a distribution depending on the parameter $\theta$. The missingness pattern encoded in the indicator matrix $\M{M}$ is then generated conditional on the data $\M{Y}$ and the parameter $\phi$. We next discuss three standard assumptions taken on the form of the conditional distribution $f(\M{M} \mid \M{Y}, \phi)$.

When $\M{M}$ is conditionally independent of the data $\M{Y}$ and $f(\M{M} \mid \M{Y}, \phi) = f(\M{M} \mid \phi)$, the data are said to be \textbf{missing completely at random (MCAR)} ~\citep{littlerubin}.  When $\M{M}$ is conditionally dependent only on the observed portion of the data $\M{Y}_{\text{obs}}$ and $f(\M{M} \mid \M{Y}, \phi) = f(\M{M} \mid \M{Y}_{\text{obs}}, \phi)$, the data are said to be \textbf{missing at random (MAR)}~\citep{littlerubin}.  When $\M{M}$ depends on the unobserved portion of the complete data and no simplifications in $f(\M{M} \mid \M{Y}, \phi)$ are possible, we say that the data are \textbf{not missing at random (NMAR)}~\citep{littlerubin}. The NMAR scenario is common in social surveys when respondents refuse to answer certain questions. For example, lower-income or higher-income respondents may decline to report income data.  Alternatively, in survey data on countries such as the World Bank Development Indicators~\citep{worldbank}, poorer countries may sometimes lack the means to collect and compile data in particular variables and so fail to report particular development indicators.  In these scenarios, the missingness depends on the value of the unobserved entries in ${\bf Y}$.

\subsection{Clustering Methods}
\label{sec:clusteringmethods}

Our numerical experiments compare methods with readily available R packages for imputation and clustering on the Comprehensive R Archive Network (CRAN) utilizing a minimum of specifications and the default settings on all imputation methods.  We also assume that the missingness mechanism in the data is unknown.

Since multiple imputation is the state-of-the-art framework for handling missing data~\citep{rubin1996, rubin1987}, we compare clusterings obtained after imputation using mainstream packages for multiple imputation in R.  These include the {\tt Amelia}, {\tt mice}, and {\tt mi} packages for multiple imputation~\citep{Amelia,mice,mi}.  When available, we employ the package functions for pooling the imputed data to combine the multiple imputations into a single dataset for use in clustering.  Otherwise, we pool the imputed data using element-wise means.  

The assumptions made in the imputations are as follows.  The {\tt Amelia} package assumes that the data are MAR.  Since our experiments employ numeric data, the default {\tt mice} settings assume that the data are MAR and specify a conditional distribution for each variable.  The default settings on the {\tt mi} package also assume that the data are MAR and specify variable-by-variable conditional models.

We obtain clusterings after imputation using the {\tt kmeans} function in the base {\tt stats} package in R \citep{r}.  In order to compare performance by deletion, we also compare $k$-means clustering after deletion of variables containing missing entries.  The algorithm for $k$-POD can be found in the {\tt kpodclustr} package for R~\citep{kpodclustr}. 

\subsection{Data}
\label{sec:datasets}
We employ one real and two simulated datasets in the numerical experiments.  

To gain a sense of how the approaches fare at extending $k$-means on real data that may not be ideally suited for the $k$-means algorithm even when data is complete, we employ the ``wine'' dataset from the UCI Machine Learning repository~\citep{uciml} containing 178 observations on 14 variables (13 represent varying chemical constituents found in each wine, and one denotes the wine classification) in three clusters. We construct 300 perturbations on the data by adding normally distributed noise with mean $0$ and standard deviation equal to one tenth of the mean value in each variable. Adding normally distributed noise with the same standard variation to each variable perturbs variables of low mean value too much and perturbs variables of high mean value too little.  Consequently, we adjust the amount of noise added to each variable to account for the different scales of variation in each variable.  

To gain a sense of how the approaches fare at extending $k$-means to missing data, we also employ two simulated datasets that $k$-means performs well on in the absence of missing data. The simulated datasets consist of mixtures of multivariate normal distributions where each normal component has identical isotropic covariances and differ only in their means. We draw a total of $500$ observations on $100$ variables from $k=10$ and $k=25$ clusters. The $k$th component has mean $\V{\mu}_k$ and covariance $\M{\Sigma} = 10\M{I}$. The means $\V{\mu}_k \in \Real^{100}$ consist of 100 i.i.d.\ draws from a 0 mean normal with standard deviation of 10. We randomly assign observations to one of $k$ cluster centroids with probability $\frac{1}{k}$. We then generate 300 realizations of data according to this design, with 100 realizations for each of the three different missingness mechanisms. 


\subsection{Experimental Setup}
\label{sec:experimentalsetup}

Table~\ref{table:simscenarios} depicts the numerical experiments by method and missingness mechanism scenario.  
\begin{table}[t!]
    \begin{center}
    \caption{Numerical experiments by method and missingness mechanism.}
    \begin{tabular}{@{}llll@{}}
    \toprule
    \multicolumn{1}{c}{$k$-Means Clustering Approaches} & \multicolumn{1}{c}{MCAR} & \multicolumn{1}{c}{MAR} & \multicolumn{1}{c}{NMAR}  \\ \midrule
    {\tt Amelia}-imputed data & \multicolumn{1}{c}{$\times$} & \multicolumn{1}{c}{$\times$} & \multicolumn{1}{c}{$\times$} \\
    {\tt mi}-imputed data & \multicolumn{1}{c}{$\times$} & \multicolumn{1}{c}{$\times$} & \multicolumn{1}{c}{$\times$}  \\
    {\tt mice}-imputed data & \multicolumn{1}{c}{$\times$} & \multicolumn{1}{c}{$\times$} & \multicolumn{1}{c}{$\times$}  \\
    $k$-POD & \multicolumn{1}{c}{$\times$} & \multicolumn{1}{c}{$\times$} & \multicolumn{1}{c}{$\times$}  \\ 
    Deleted data &  & \multicolumn{1}{c}{$\times$} &    \\ \bottomrule
    \end{tabular}
        \label{table:simscenarios}
    \end{center}
\end{table}
To simulate the MCAR mechanism, we randomly remove entries to obtain approximately 5, 15, 25, 35, and 45 percent overall missingness in the wine dataset and 25, 50, and 75 overall missingness in the simulated datasets.  To simulate the MAR mechanism, we randomly remove entry values in the 1st, 4th, and 7th columns in the wine dataset to obtain a spread of overall missingness.  We did not simulate the MAR mechanism in experiments utilizing simulated data due to the impracticality of obtaining overall missingness levels of 25, 50, and 75 percent when missing entries are restricted to a subset of the columns.  To simulate the NMAR mechanism, we randomly remove entries in approximately the bottom 5th, 15th, 25th, 35th, or 45th quantiles in each of the variables of variables in the wine dataset, and in the 25th, 50th, and 75th quantiles in the simulated dataset.

Higher levels of missingness --- such as 75 percent missingness --- are traditionally uncommon.  The recent explosion of large and sparse data from sources such as online social networks and recommender systems, however, has resulted in more data with higher levels of missingness.  For example, the Netflix Prize competition data had nearly 99 percent missingness (approximately 100 million ratings from 500,000 users on 18,000 movies)~\citep{netflixprize}.  Accompanying this trend towards large and sparse data is an increasing demand for methods that can handle data with such large missingness percentages.  The experiments with 75 percent missingness offer a sense of how different approaches might perform on these kinds of data.

The clustering experiments comprise 100 trials per combination of clustering method, missingness mechanism, and missingness percentage.  The MCAR and NMAR scenarios exclude deletion since missing entries appear in all variables.

We employ the Rand score obtained by comparing each clustering result to the true class label variable as the comparison metric for clustering results.  The Rand score is a commonly used metric for quantifying the agreement between two partitions of a finite set \citep{Ran1971}, and ranges between $0$ for no agreement and $1$ for perfect agreement.  Higher scores indicate greater similarity with the true class labels and hence, more accurate clustering performance.  We utilize the {\tt adjustedRand} function in the {\tt clues} package for R to compute Rand scores~\citep{clues}.

We do not calculate aggregate statistics on Rand scores for clustering approaches when they fail to complete experiments.  Failure to complete occurs when approaches fail to produce clustering results for all simulations in a given scenario.  Since all approaches other than $k$-POD employ the $k$-means algorithm to produce a clustering result, inability to complete experiments occurs when methods do not complete an imputation.  In experiments using the simulated datasets, we also terminate experiments for each scenario after 55 hours.

To compare timing results, we record the time (in seconds) required to obtain a clustering.  We perform computations in serial on a multi-core computer with 24 3.3 GHz Intel Xeon processors and 189 GB of RAM.  Since $k$-means is sensitive to scale, we scale each partially observed dataset prior to clustering using the default {\tt scale} function in R.

\subsection{Results}
\label{sec:results}

The following tables depict the results of the numerical experiments.  An empty space in the table indicates failure to complete the experiment for a given scenario.  We report actual missingness percentages in the columns.

Table~\ref{fig:winerand} reports the mean and standard error of the Rand scores in experiments utilizing the wine dataset.  At low levels of missingness, all methods perform well regardless of the missingness mechanism.  When the data are MAR and missing in only a few variables, even deletion yields reasonable results.  As the percentage of missingness increases, however, some approaches fail to complete experiments.  For instance, {\tt Amelia} did not complete imputations beginning at 35, 16, and 36 percent missingness in the MCAR, MAR, and NMAR scenarios, respectively.  Similarly, {\tt mice} did not complete imputations starting at 19 percent missingness in the MAR scenario.

In contrast, $k$-POD produces the most accurate results in scenarios beginning at 45 and 12 percent overall missingness in the MCAR and MAR scenarios, respectively.  Moreover, even when other approaches produce more accurate results, the $k$-POD clusterings are very comparable.

	\begin{table}[th]
	   \caption{Clustering results for experiments utilizing the wine dataset by mechanism, method, and unobserved percentage (3 clusters).}
	    \begin{tabular}{lcl}
		\includegraphics[width=1.0 \textwidth]{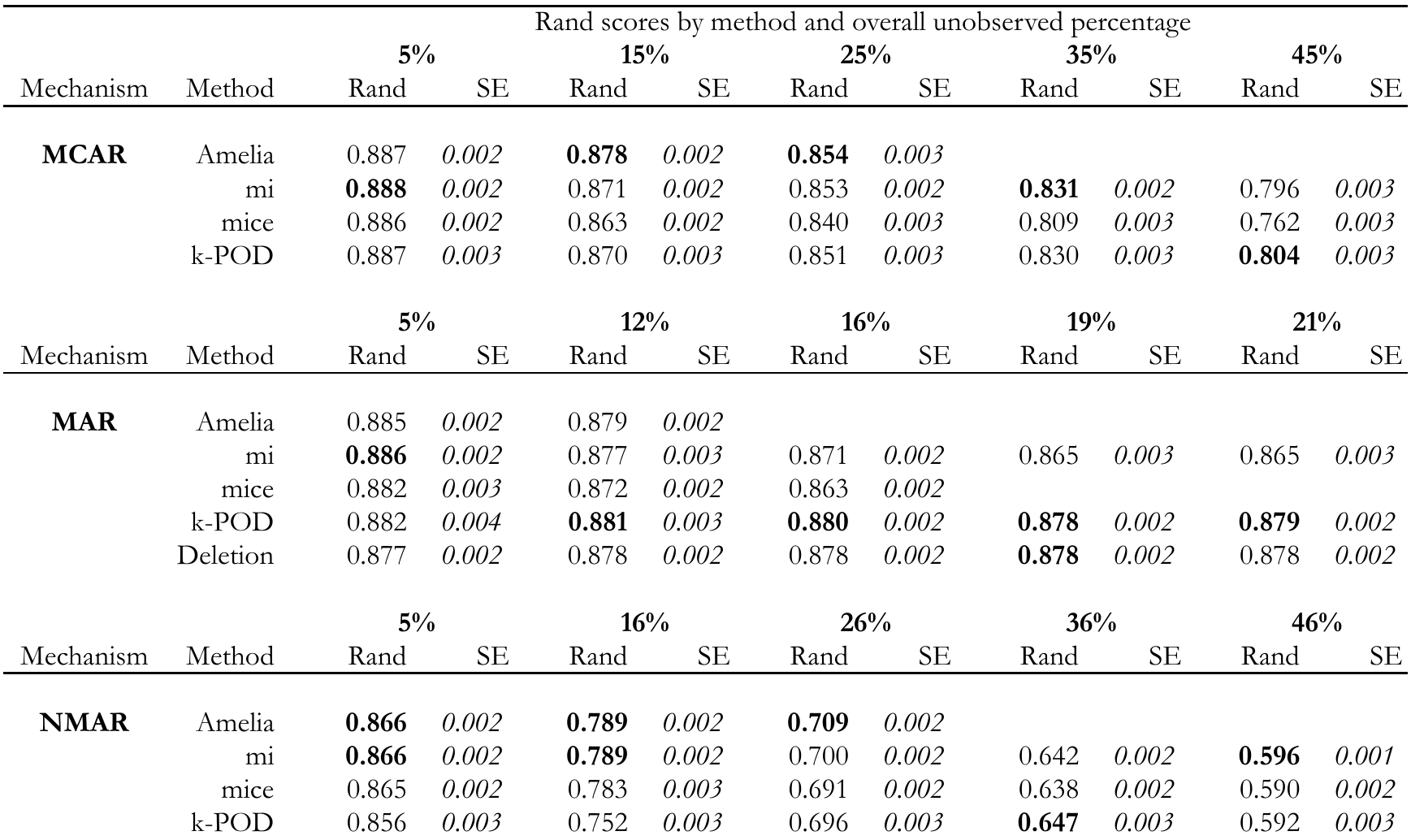}
	     \end{tabular}
	     \label{fig:winerand}
	\end{table}

Table~\ref{fig:winetiming} reports the mean and standard error of the time (in seconds) required to compute a clustering result for the wine dataset.  It shows that {\tt mi} is the slowest in all scenarios and in the MAR scenarios, deletion is very expeditious.

It also shows that $k$-POD is essentially as fast as {\tt Amelia}, appreciably faster than {\tt mice} in the MCAR and NMAR scenarios, and significantly faster than {\tt mi} in all scenarios.  Notably, although {\tt Amelia} is fast when it is able to obtain imputations, it fails at higher levels of overall missingness.  Additionally, {\tt mi} is the only imputation-based approach to complete all experiments, regardless of missingness mechanism or overall missingness percentage.  However, it requires more computational time compared with other methods, particularly when the missingness occurs in all variables.

	\begin{table}[th]
		\caption{Timing results for experiments utilizing the wine dataset by mechanism, method, and unobserved percentage (3 clusters).}
	    \begin{tabular}{lcl}
		\includegraphics[width=1.0 \textwidth]{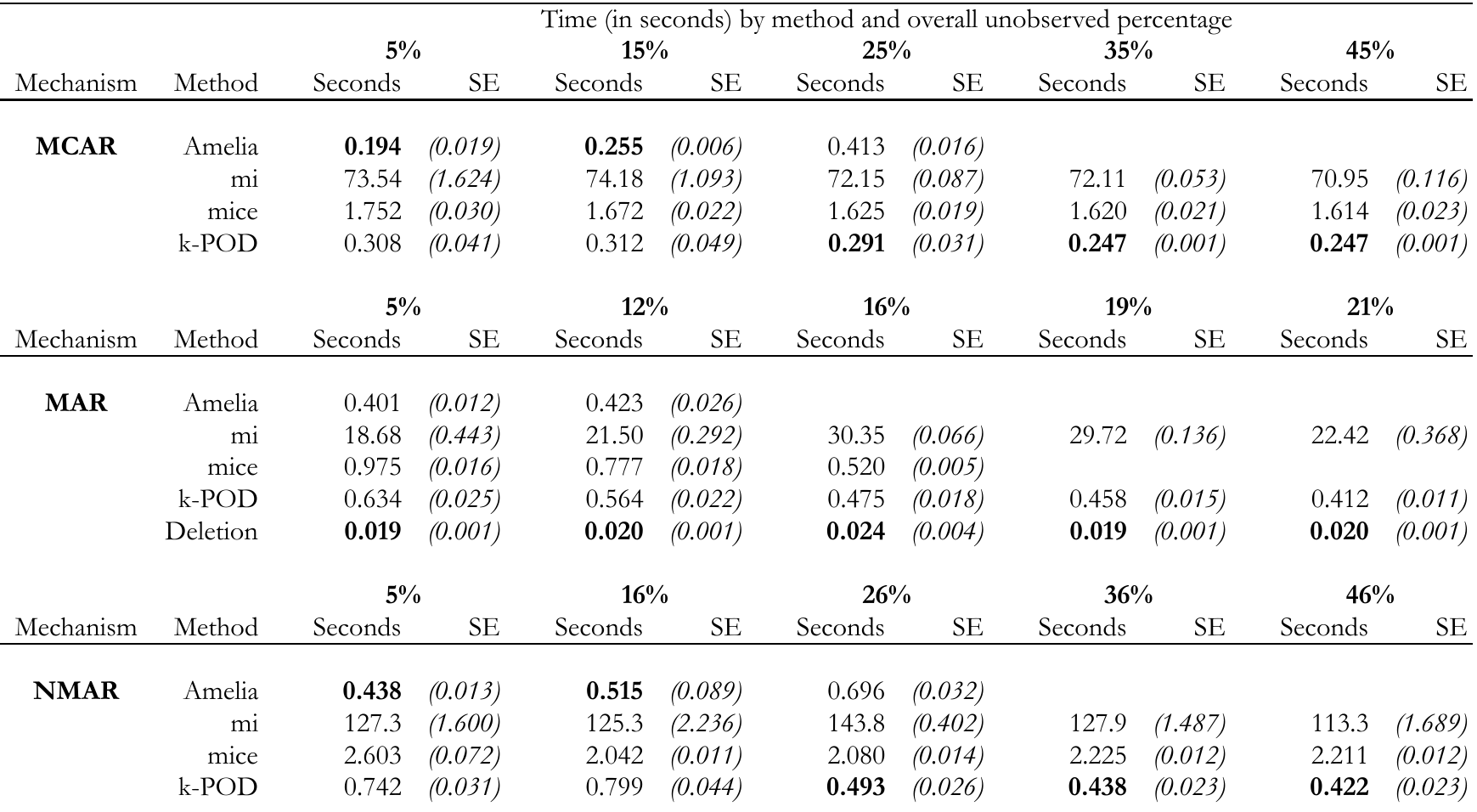}
	     \end{tabular}
    		\label{fig:winetiming}
	\end{table}

Table~\ref{fig:simk10rand} reports the mean and standard error of the Rand scores in numerical experiments using simulated data with $10$ clusters.  The table shows that {\tt Amelia} and {\tt mi} did not produce results within 55 hours.  Overall, $k$-POD produces the most accurate results at 75 and 50 percent missingness in the MCAR and NMAR scenarios, respectively, and produces results that are essentially equivalent to those obtained using {\tt mice} in the remaining experiments.  The timing results in Table~\ref{fig:simk10timing} show that {\tt mice} requires significantly more computation time than $k$-POD to produce essentially the same results.

	\begin{table}[th]
		\caption{Clustering results for experiments utilizing the simulated dataset by mechanism, method, and unobserved percentage (10 clusters).}
		\centering
	    \begin{tabular}{lcl}
		\includegraphics[width=0.8 \textwidth]{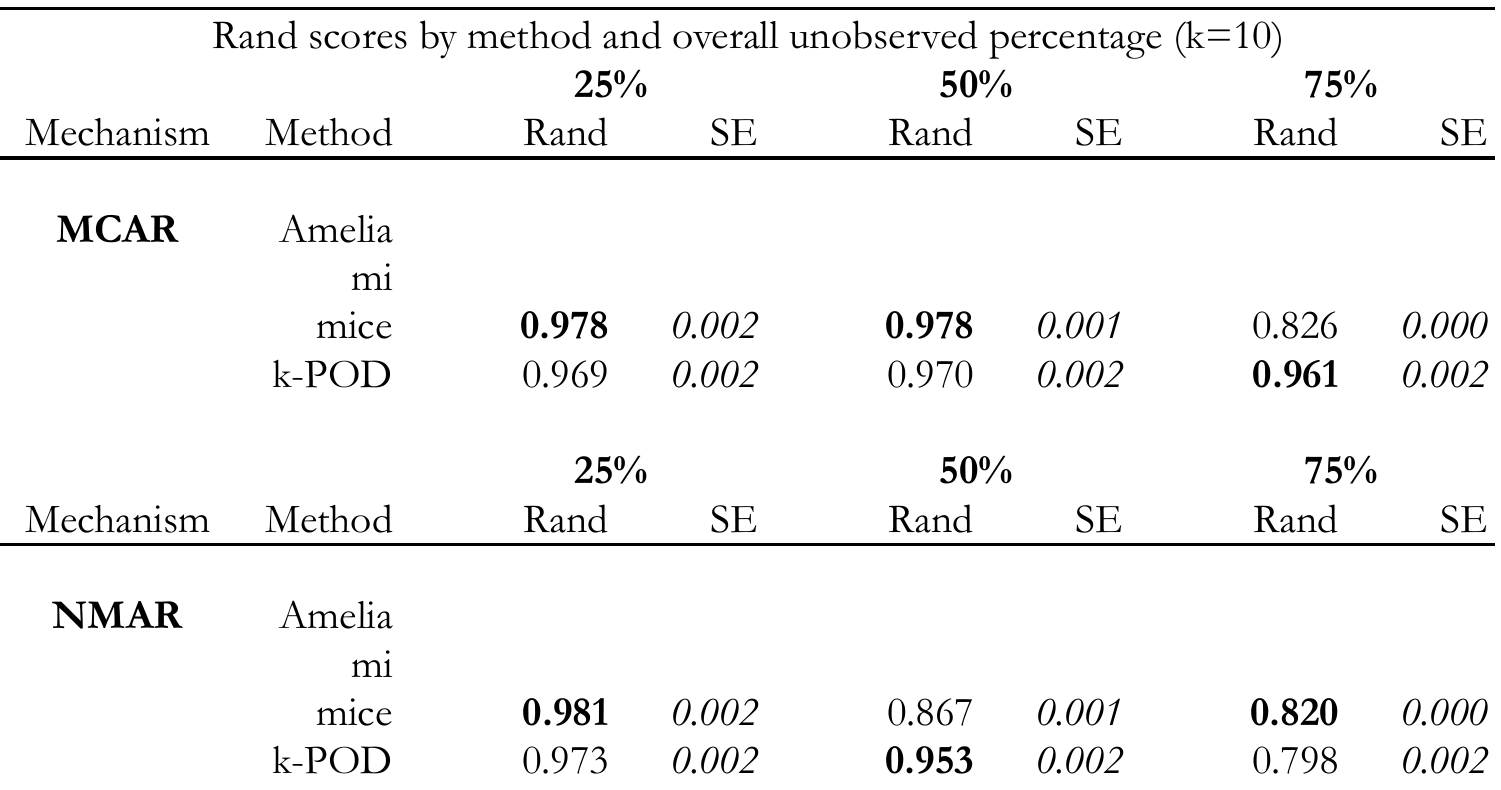}
	     \end{tabular}
    		\label{fig:simk10rand}
	\end{table}

	\begin{table}[th]
		\caption{Timing results for experiments utilizing the simulated dataset by mechanism, method, and unobserved percentage (10 clusters).}
		\centering
	    \begin{tabular}{lcl}
		\includegraphics[width=0.8 \textwidth]{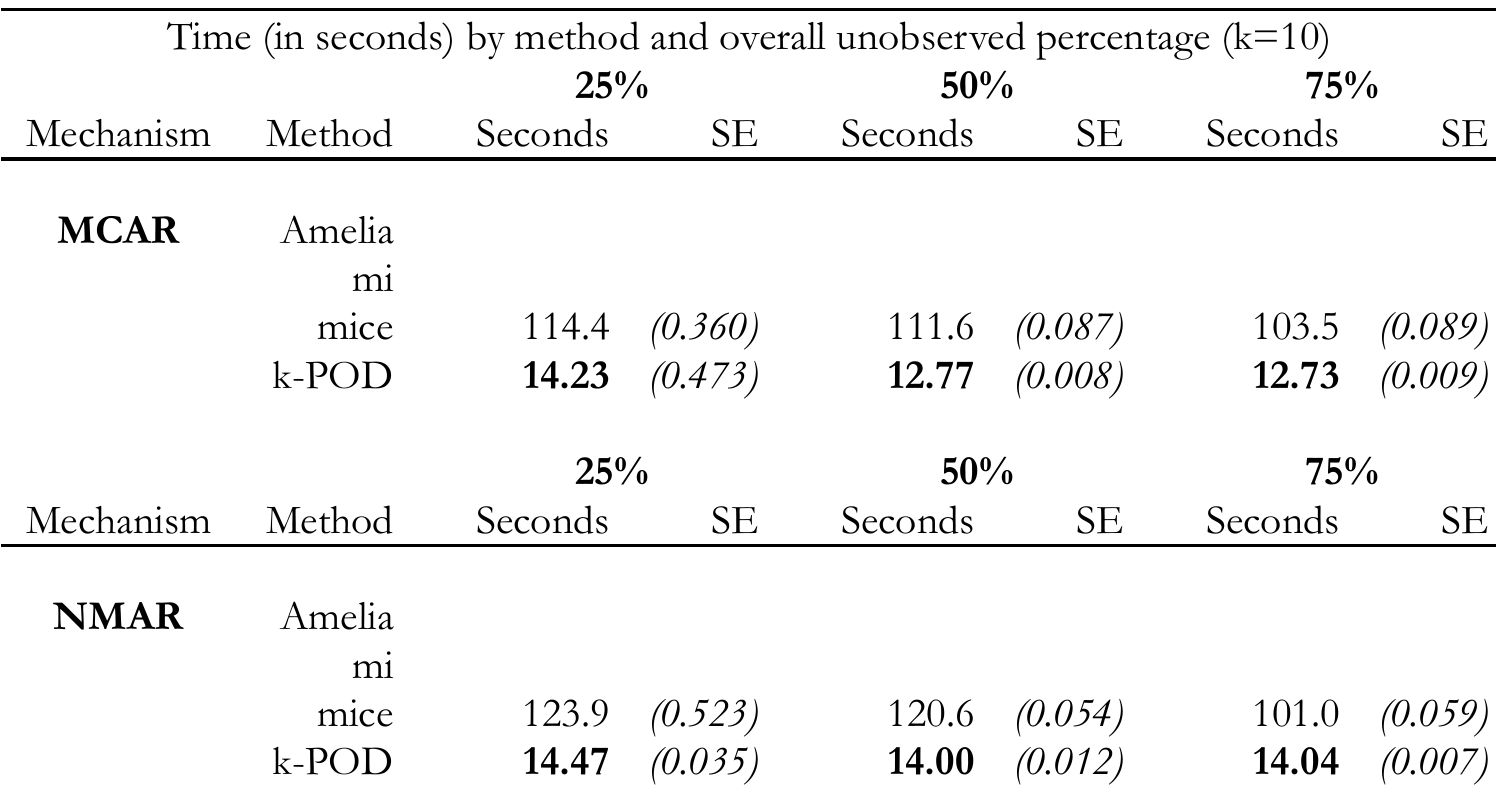}
	     \end{tabular}
    		\label{fig:simk10timing}
	\end{table}

Table~\ref{fig:simk25rand} reports the mean and standard error of the Rand scores in experiments using simulated data with $25$ clusters. Again, {\tt Amelia} and {\tt mi} did not produce results within 55 hours.  While the results between $k$-POD and {\tt mice} remain similar, $k$-POD produces the most accurate results in all scenarios except one.  The timing results in Table~\ref{fig:simk25timing} again show that, in general, {\tt mice} requires significantly more computation time than $k$-POD to produce essentially the same results.  When the data are MCAR and there are many clusters and low overall missingness, however, $k$-POD takes approximately as long as {\tt mice} to converge to a clustering result.  The standard error in the $k$-POD timing result for the $k=25$ MCAR scenario with 25 percent missingness suggests that in those scenarios, $k$-POD may take slightly longer to converge to a clustering result that is in accord with the observed data depending on its initialization.  Notably, $k$-POD produces accurate results, regardless of the initialization.  Finally, the timing results for $k=25$ clusters suggest that $k$-POD may require more computational time for larger numbers of clusters.  In contrast, the timing results for {\tt mice} were very comparable between $k=10$ and $k=25$ clusters.

	\begin{table}[t]
		\caption{Clustering results for experiments utilizing the simulated dataset by mechanism, method, and unobserved percentage (25 clusters).}
			    \centering
	    \begin{tabular}{lcl}
		\includegraphics[width=0.8 \textwidth]{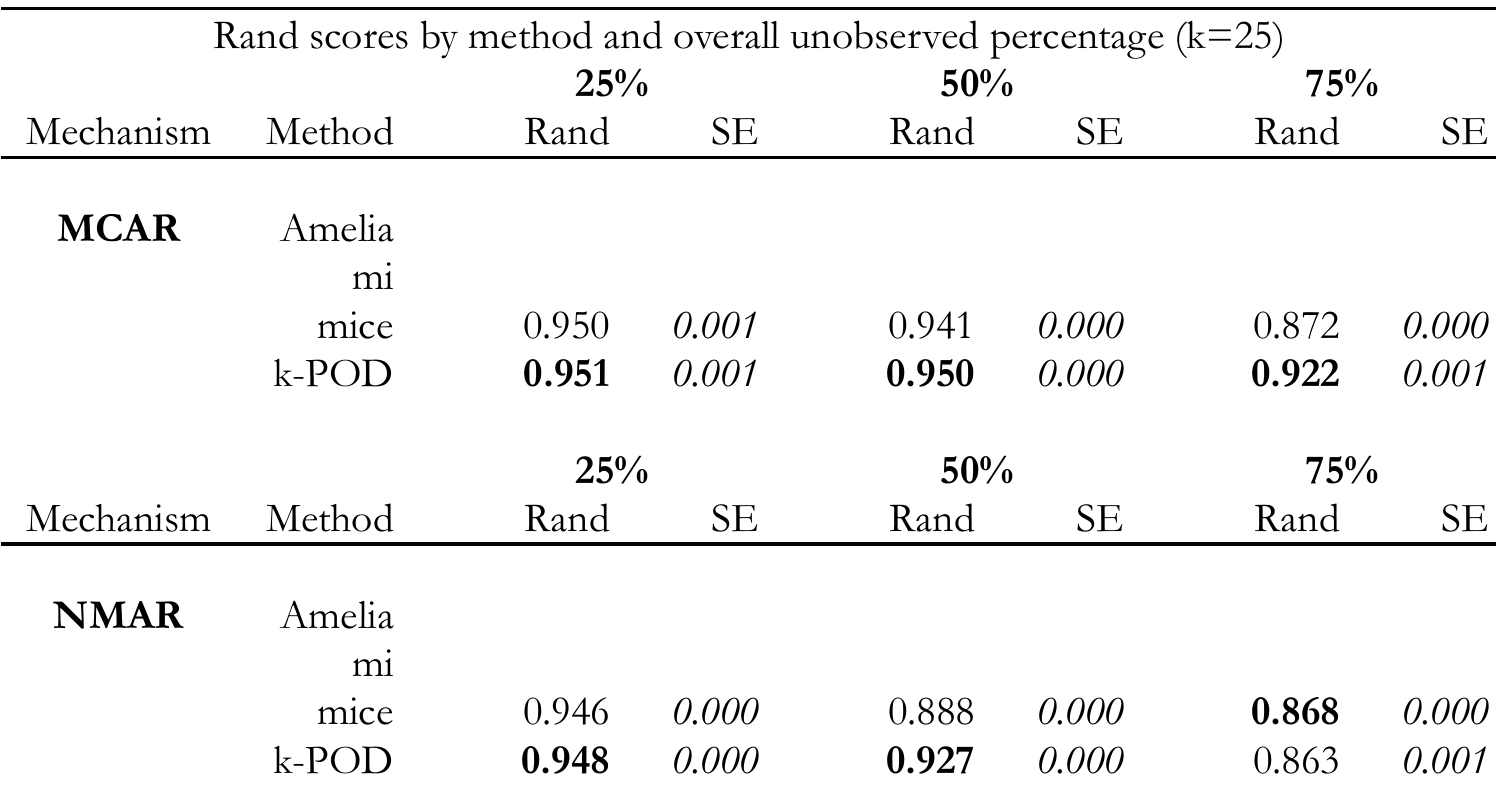}
	     \end{tabular}
    		\label{fig:simk25rand}
	\end{table}

	\begin{table}[t]
		\caption{Timing results for experiments utilizing the simulated dataset by mechanism, method, and unobserved percentage (25 clusters).}
		\centering
	    \begin{tabular}{lcl}
		\includegraphics[width=0.8 \textwidth]{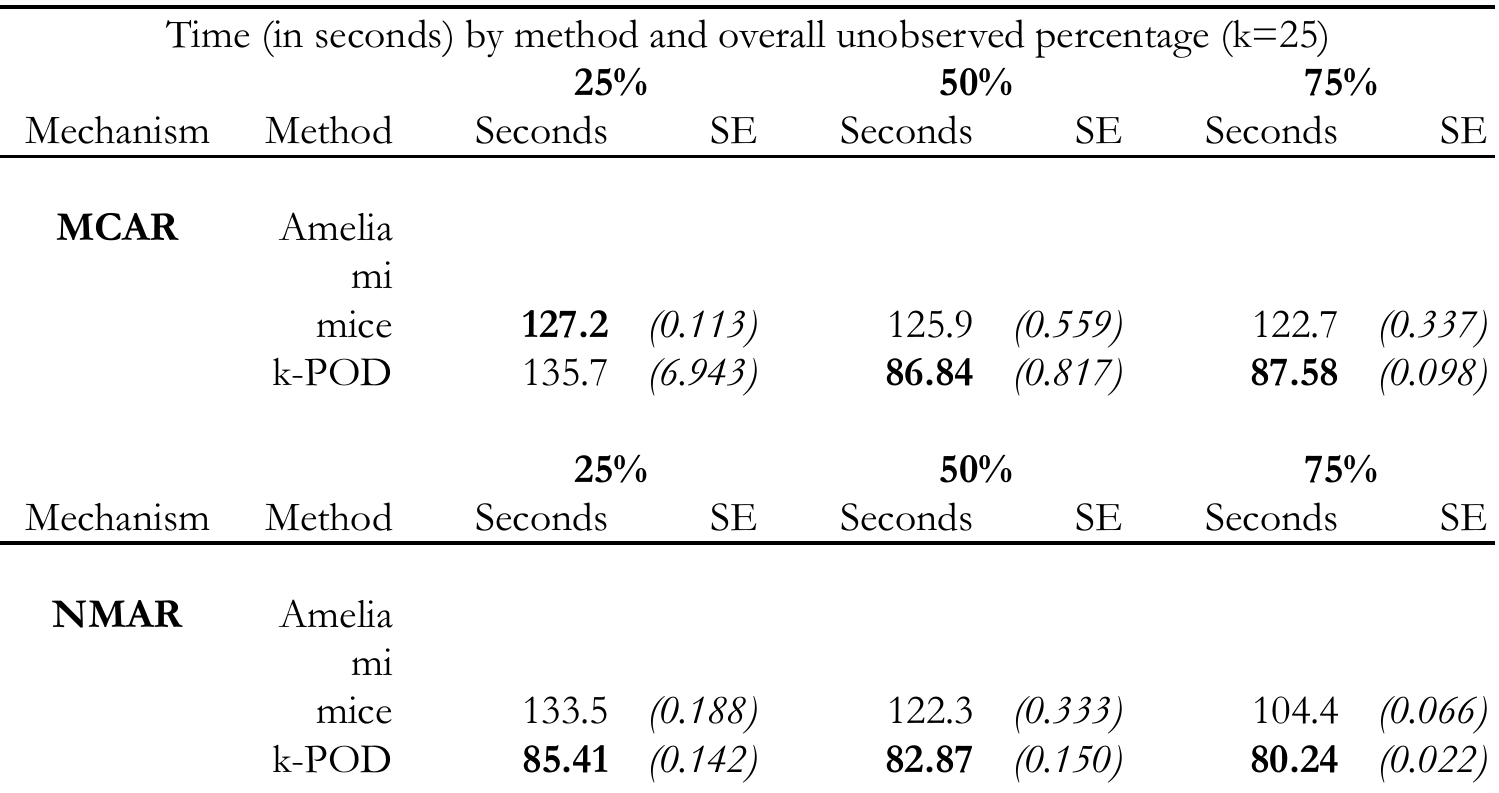}
	     \end{tabular}
    		\label{fig:simk25timing}
	\end{table}

The experiments with the simulated data highlight how imputation can become prohibitively expensive on larger data.  Even in scenarios with 500 observations on 100 variables, some imputation-based approaches did not produce results within 55 hours.

\clearpage

\section{Discussion}
\label{sec:discussion}

The $k$-POD method offers a simple, reliable, and fast alternative to deletion and imputation in clustering missing data.
Two key facts enable $k$-POD to identify a clustering using no less, and no more, than the observed data. The first is that the $k$-means problem can be formulated as seeking an optimal rank $k$ decomposition of the data matrix. Once in this form, the residual sum of squares over the observed values is a natural way to formulate a missing data version of the $k$-means loss. The second is that this new loss admits a simple majorization that can be inexactly minimized with the $k$-means algorithm.

The numerical results demonstrate that $k$-POD is not only accurate, but also fast --- particularly at higher levels of overall missingness.  There are two reasons for this.  First, the majorization step consists of simply copying the relevant entries of the centroid estimates from the $k$-means step into the missing entries in the data matrix.  Second, the minimization step consists of running $k$-means, which is fast in itself; indeed, each iteration requires $\mathcal{O}(knp)$ computations, which is linear in the data.  Particularly when moving to larger data, the setup and computational costs required to obtain reasonable imputations may become prohibitively expensive, as exhibited in the experiments with 500 observations on 100 variables.

Of course, $k$-POD is not a panacea for clustering partially observed data.  As a meta-algorithm built around the $k$-means algorithm, it retains the limitations common to $k$-means clustering.  For example, in the complete data case, $k$-means tends to struggle when clusters overlap or when the scatter within-clusters varies drastically from cluster to cluster. In these cases, when the data are completely observed, EM clustering can produce superior results.  Analogously, in these cases, when data are missing, the more complicated mixture model extensions mentioned at the end of the introduction may be required.

Nonetheless, the simple and efficient $k$-means algorithm, despite its shortcomings and the availability of more sophisticated alternatives, remains the often-preferred choice for clustering data.  Its ubiquity warrants an extension for handling missing data that maintains its simplicity and efficiency.  The $k$-POD method builds upon $k$-means clustering to provide a simple and quick alternative to clustering missing data that works even when the missingness mechanism is unknown, when external information is unavailable, and when there is significant missingness in the data.

The software for implementing $k$-POD can be found in the {\tt kpodclustr} package for R and is available on CRAN \citep{kpodclustr}.

\section{Acknowledgements}

This material is based upon work supported by, or in part by, the U. S. Army Research Laboratory and the U. S. Army Research Office under contract/grant number ARO MURI W911NF0910383.

\bibliographystyle{asa}
\bibliography{kPOD}

\end{document}